\font\BBig=cmr10 scaled\magstep2

\input epsf

\def\title{
{\bf\BBig
\centerline{
Vortices in the Landau--Ginzburg Model}
\medskip
\centerline{of the}
\medskip
\centerline{Quantized Hall Effect}
\bigskip
}
} 


\def\authors{
\centerline{
M. HASSA\"INE\foot{e-mail: hassaine@cecs.cl},
P.~A.~HORV\'ATHY\foot{e-mail: horvathy@univ-tours.fr}
 and
J.-C.~YERA\foot{e-mail: yera@univ-tours.fr}}
\medskip
\centerline{
D\'epartement de Math\'ematiques}
\medskip
\centerline{Universit\'e de Tours}
\medskip
\centerline{Parc de Grandmont,
F--37200 TOURS (France)
}
}

\def\runningauthors{
Hassa\"\i ne, Horv\'athy \&
Yera
}

\def\runningtitle{
Vortices in the Landau--Ginzburg model of the QHE
}


\voffset = 1cm 
\baselineskip = 16pt 

\headline ={
\ifnum\pageno=1\hfill
\else\ifodd\pageno\hfil\tenit\runningtitle\hfil\tenrm\folio
\else\tenrm\folio\hfil\tenit\runningauthors\hfil
\fi
\fi}

\nopagenumbers
\footline={\hfil} 


\def\and{\qquad\hbox{and}\qquad}

\def\kikezd{\parag\underbar} 

\def\smallover#1/#2{\hbox{$\textstyle{#1\over#2}$}}
\def\2{{\smallover 1/2}}
\def\ccr{\cr\noalign{\medskip}} 
\def\parag{\hfil\break} 
\def\={\!=\!}
\def\p{\partial}

\def\L{{\cal L}}

\def\vn{\mathop{\vec{\nabla}}}
\def\L{{\cal L}}


\newcount\ch 
\newcount\eq 
\newcount\foo 
\newcount\ref 

\def\chapter#1{
\parag\eq = 1\advance\ch by 1{\bf\the\ch.\enskip#1}
}

\def\equation{
\leqno(\the\eq)\global\advance\eq by 1
}

\def\eqletter#1{
\leqno(\the\eq{#1})
}

\def\foot#1{
\footnote{($^{\the\foo}$)}{#1}\advance\foo by 1
} 

\def\reference{
\parag 
}

\ch = 0 
\foo = 1 
\ref = 1 
\eq = 1


\title
 
\authors

\parag{\bf Abstract.}
{\it The  `Landau--Ginzburg' theory
of Girvin and MacDonald, 
modified by adding the natural magnetic term, 
is shown to admit stable topological as well as
non-topological vortex solutions.  
The system is the commun $\lambda\to0$ limit of
two slightly different non-relativistic 
Maxwell--Chern--Simons 
models of the type introduced recently by Manton.
The equivalence with the model of Zhang, Hansson and 
Kivelson is demonstrated.}

\bigskip\medskip
\noindent
{\sl Journal of Physics} {\bf A} (Math. Gen.) {\bf 31}, 9073-79 (1998)
\bigskip\medskip


In Ref.~[GIR], Girvin and MacDonald presented a ``Landau--Ginzburg'' 
theory for the Quantum Hall Effect.  
On phenomenological grounds, they suggest to represent the
off-diagonal long range order (ODLRO) by a scalar field $\psi(\vec{x})$ 
on the plane,
and the frustration due to deviations away from the quantized Laughlin 
density by
an effective gauge potential $\vec{a}(\vec{x})$.  
We propose to describe this 
static planar system by the Lagrange density
$$
\L
=
{1\over2}b^2 +\Bigl| \vec{D}\psi\Bigr|^2 
+i\phi\bigl(|\psi|^2-1\bigr)
-i{\kappa\over2}\Bigl(\phi\vn\times\vec{a}
+\vec{a}\times\vn\phi
\Bigr),
\equation
$$
where $b=\!\vn\times\vec{a}$ is the effective magnetic field,
$\vec{D}=\!\vn +i\,\vec{a}$ 
is the gauge-covariant derivative, and the Lagrange multiplier $\phi$ 
is a scalar potential. Eq.~(1) only
differs from the original expression of Girvin and MacDonald in our having 
added the natural magnetic term, ${b^2/2}$, also
present in conventional Landau-Ginzburg theory [LP].  
The associated equations of motion read
$$
\vec{D}^2\psi=i\phi\,\psi, 
\eqletter{a}
$$
\null\vskip-11mm
$$
\kappa b=|\psi|^2-1, 
\eqletter{b}
$$
\null\vskip-10mm
$$
\vn\times b-i\kappa\vn\times\phi=-\vec{\jmath}, 
\eqletter{c}
$$
where 
$\vec{\jmath}=-i\big(\psi^*\vec{D}\psi-\psi(\vec{D}\psi)^*\big)$
 the current.
The first is a static, gauged Schr\"odinger equation.  
The second is the
relation proposed by Girvin and MacDonald to relate the magnetic 
field to the particle density. 
Note here the $-1$ coming from the
weird term $-i\phi$ in the Lagrangian,
and representing the background charge [MAN].
The last equation is the Amp\`ere--Hall law~: 
$\vec{e}=-i\vn\phi$ is an effective electric field, so that
$\kappa$ is interpreted as the {\it Hall conductance}.

This system is rather similar to those studied in Chern-Simons field 
theory [JP],
and in particular to that recently introduced by Manton [MAN], [HHY].
Using these techniques, (i) we construct, with the method of 
Bogomolny [BOG], stable vortex solutions;
(ii) point out that (1) is the $\lambda\to0$ limit of two
slightly different systems;
(iii) demonstrate the equivalence with another
`Landau-Ginzburg' model, introduced by Zhang {\it et al. \/} [ZH]. 

Let us try in fact to reduce the second-order equations to the 
first-order ``self-dual'' system
\advance\eq by 1
$$
(D_1 \pm iD_2)\psi=0, 
\qquad
\kappa b=|\psi|^2-1.
\equation
$$
From the first of these relations we infer that $\vec{D}^2=\pm b$ and 
$\vec{\jmath}=\mp\vn\times\varrho$. 
 Inserting into the Schr\"odinger equation
determines the multiplier field as 
$
\phi=\mp{i\over\kappa}(\varrho-1).
$
Then, from Amp\`ere's law we get that the Hall conductance $\kappa$ has 
to be
$$
\kappa=\pm{1\over2}.
\equation
$$
The vector potential is expressed using the self-dual An\-satz (3) as
$\vec{a}=\mp(1/2)\vn\times\log\varrho+\vn\omega$,
 where $\omega$ is an
arbitrary real function chosen so that $\vec{a}$ is regular [JP].  
Inserting this into (2b) 
we end up, for both signs, with the ``Liouville-type'' equation
$$
\Delta\log\varrho=4(\rho-1).
\equation
$$
Note that  any solution will carry an effective  magnetic as well as 
an electric field.

\goodbreak
Now we have to tell what kind of solutions we are interested in.  
To see this, let us consider the energy,
$$
H=\int\biggl\{ 
{b^2\over2}+|\vec{D}\psi|^2 \biggr\}d^2\vec{x}.
\equation
$$
Using the identity 
$|\vec{D}\psi|^2=|(D_1\pm iD_2)\psi|^2\mp b|\psi|^2 
+{\rm(surface\ term)}$\ ,
 as well as (2b), the energy is rewritten as
$$
H=\int\biggl\{|(D_1\pm iD_2)\psi|^2
+\Bigl({1\over2\kappa^2}\mp{1\over\kappa }\Bigr)|\psi|^4
+{1\over2\kappa^2}
+\Bigl(-{1\over\kappa^2}\pm{1\over\kappa}\Bigr) 
|\psi|^2 \biggr\}d^2\vec{x}.
$$
The quartic term disappears when $\kappa=\pm{1/2}$, 
leaving us with
$$
H=\int\biggl\{|(D_1\pm iD_2)\psi|^2+2(1-|\psi|^2)\biggr\}d^2\vec{x}.
\equation
$$
The system admits the zero-energy ground state (condensate) 
$\psi\equiv 1$ and $\vec{e}=0,\ b=0$.  

Then we have two possibilities:

 $\bullet$ Either, to
get finite-energy, we can require that at infinity the solution tends to 
the condensate state, $|\psi|\to1$ and $|\vec{D}\psi|\to0$. 
These two conditions imply that
the magnetic flux is quantized,
$$
\Phi\equiv
\int b\, d^2x=2\pi n,
\equation
$$
where the integer $n$ is the `winding number' of $\psi$, which maps the 
circle at infinity into $U(1)$.  Then, using Eq.~(2b), 
the `particle number'
$$
N=\int\big(1-|\psi|^2\big)\,d^2\vec{x}
\equation
$$
is finite, and is related to the flux as 
$N=-\kappa\Phi=-2\pi\kappa n$.  
($N$ is conserved owing to the continuity equation
which follows from Eq. (2a).) 
Our finite-energy solutions 
(referred to as  {\it topological vortices}) carry hence
a non-vanishing flux as well as a charge. 
Their energy satisfies, by Eq.~(7), the `Bogomolny' inequality
$$
H\ge 2N=-4\pi\kappa n=\mp2\pi n.
\equation
$$
To get a positive bound, $\kappa$ and $n$ must have opposite 
signs~: the upper (lower) sign works for
$n<0$ (resp. for $n>0$).

The `Bogomolny' bound (10)
is saturated when the self-duality eqns.~(3) hold.  
(This is in fact the case of `Bogomolny' vortices in 
the Abelian Higgs model, so that
Eq.~(5) admits a $2n$-parameter family of solutions [WEIN]). 
Since they correspond to the absolute minima of the energy,
such solutions are stable.

\goodbreak
$\bullet$ Another possibility is, however, to choose 
$b=b_0=-{1/\kappa},\ \vec{e}=0,\ \psi\equiv 0$
as the ground state.  This is {\it not} a finite-energy
solution, though subtracting its (constant) energy density 
${1/2\kappa^2}=2$
from (7), we get
\eq=7
$$
\overline{H}=
\int\biggl\{{b^2\over2}-{b_0^2\over2}+|\vec{D}\psi|^2\biggr\} 
d^2x
=
\int \biggl\{|(D_1\pm iD_2)\psi|^2-2|\psi|^2 \biggr\}d^2x.
\eqletter{'}
$$

Now we can look for {\it non-topological} solutions, {\it i.e.\/} such 
whose `number', defined as
\eq=9
$$
\overline{N}=\int |\psi|^2\, d^2x,
\eqletter{'}
$$
converges. This `renormalized number' $\overline{N}$ is now
a continuous, rather than quantized parameter, which takes any positive value.
Then we get the modified Bogomolny bound
\eq=10
$$
\overline{H}\ge-2\overline{N},
\eqletter{'}
$$
with the equality attained when the self-duality eqns. (3) \ hold.
(Remember that $\overline{H}$ is the {\it relative} energy
with respect to the infinite-energy background).

Since $\psi\to0$ at infinity, the condition $D\psi\to0$ does not now 
imply a quantized flux. The integral (8) is indeed infinite, as one 
sees directly in the radial case from Eqn. (12) below. However, subtracting the 
constant background magnetic field, $b=b_0=-{1/\kappa}$, 
we get  the renormalized number $\overline{N}$ in 
Eqn. (9'):
\eq=8
$$
\overline{\Phi}\equiv 
\int\overline{b}d^2x\equiv \int \big(b-b_0\big)d^2x
=\int \smallover1/\kappa |\psi|^2\, d^2x
={1\over\kappa}\overline{N}.
\eqletter{'}
$$
This ``renormalized flux'' depends hence on a continuous parameter, 
namely on $\overline{N}$, just like for relativistic non-topological 
solitons [JLW].

The most convenient way of studying the solutions is to
work directly with the {\it first-order equations}
(3) rather then with the second-order Eqn. (5).
Assuming that the fields have the form
$\psi=f(r)e^{in\theta},\ a_r=0,\ a_\theta=a(r)$, 
the SD equations read
\eq=11
$$
f'=\pm{n+a\over r}f,
\qquad
{a'\over r}=\pm2\big(f^2-1\big).
\equation
$$
Regularity at the origin requires $n$ and $\kappa$ to be 
correlated as
${\rm sign\ } n=-{\rm sign\ }\kappa$, so that we get 
{\it chiral solitons}.
The small-$r$ behavior is
$f(r)=\alpha r^{\vert n\vert},
\ a=\mp r^{2}$, where $\alpha$ is a real parameter. 
For large $r$, we find instead
$$\matrix{
\matrix{
f(r)&\sim\hfill
&1-CK_{0}(2r)\hfill
\sim1-C\,\displaystyle{e^{-2r}\over\sqrt{r}}\hfill
\ccr
b&\sim\hfill
&D\,rK_{1}(2r)\hfill
\sim D\,\displaystyle{e^{-2r}\over\sqrt{r}}
\hfill
\cr}
\qquad\hfill
&\hbox{for a topological vortex}\hfill
\ccr\cr
\matrix{
f(r)&\sim\hfill
&e^{-r^2}\hfill
\ccr
b&\sim\hfill
&\mp2\big(1-e^{-2r^{2}}\big)
\hfill
\cr}
\qquad\hfill
&\hbox{for a non-topological vortex}\hfill
\cr}
\equation
$$

A simple numerical calculation shows that, for each integer value of 
$n$, there is just one radially symmetric topological vortex
obtained for $\alpha=\alpha_0(n)$, while
non-topological vortices arise for an entire range 
$\alpha<\alpha_0(n)$ of the parameter.
This behaviour is understood by looking,
following Ezawa {\it et al.\/} [EZA],
at the second-order equation (5).  
Again restricting ourselves to the radial 
case, we can view $x=\log f$ 
as the ``position'' and the radial coordinate, $t\equiv r>0$, 
as ``time'', so that Eq.~(5) becomes
$$
\ddot{x}
=
-{1\over t}\dot{x}-\vn U,
\qquad U(x)\equiv(2x-e^{2x}).
\equation
$$
This is the equation of motion of a classical particle 
in a time-dependent
frictional force and an external potential $U(x)$.
Observe that $U(x)$ increases 
from $x=-\infty$, reaches its maximum at $x=0$, 
after which it decreases.
As $t\to\infty$, $f\to1$ i.e. $x\to0$ for a topological soliton,
and $f\to0$ i.e. $x\to-\infty$ for a non-topological soliton.
 The regularity of $\psi$ requires $f$ to vanish at the 
origin.   
Let us hence consider a ``particle" which starts at `time'
$t=0$ from the `position' $x=-\infty$. 
If its ``energy'' is not
sufficient to climb the potential hill,
it will, after some time, fall back to $x=-\infty$ as $t\to\infty$~:
 we get a non-topological soliton. 
Increasing the initial velocity, we can make  our particle to
approach $x=0$ when $t\to\infty$, yielding a
topological soliton. 
 Clearly, this only happens for some specific initial ``energy'',
corresponding to a specitif value $\alpha=\alpha_{0}(n)$ of the initial 
parameter.
If the ``energy'' is even higher, the particle overshoots: 
the required boundary conditions can not be satisfied,
so that no self-dual vortex can exist.
 
\goodbreak
\vskip-3.3truecm
\centerline{\epsfxsize=7.8truecm\epsfbox{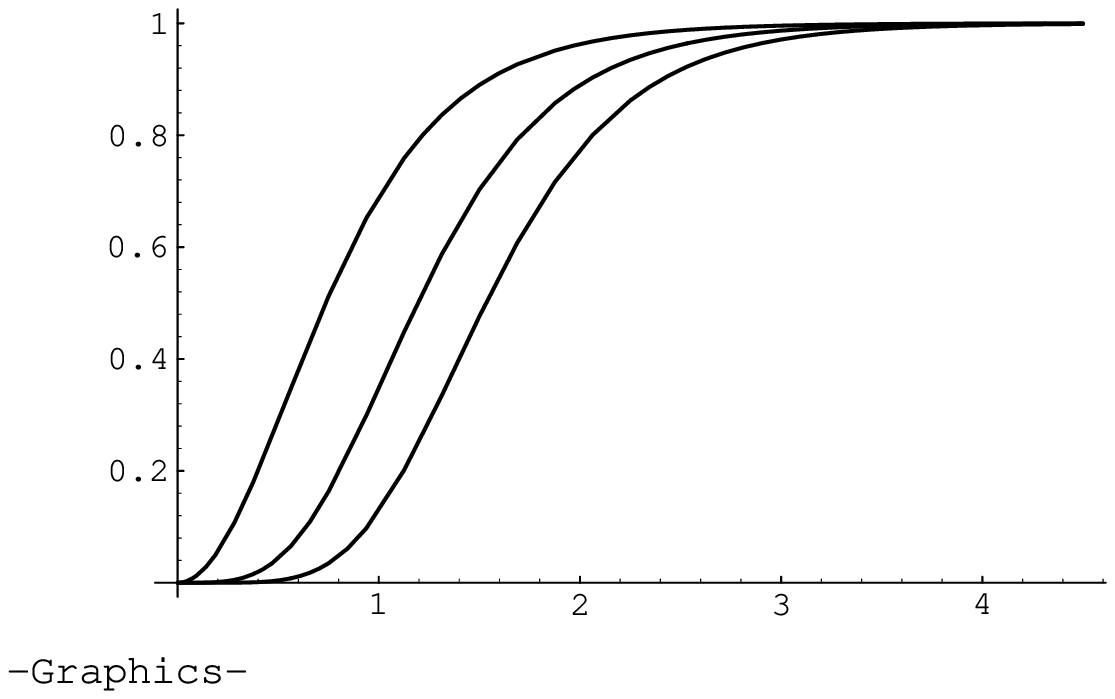}}
\vskip-0.6truecm

\kikezd{Fig. 1.}\hskip2mm
{\it The order parameter density of the 
radially symmetric topological vortices
with winding number $n=1, 2, 3$. 
For each value of $n$, there is exactly one vortex}.
\null
\vskip-3.3truecm
\centerline{\epsfxsize=7.8truecm\epsfbox{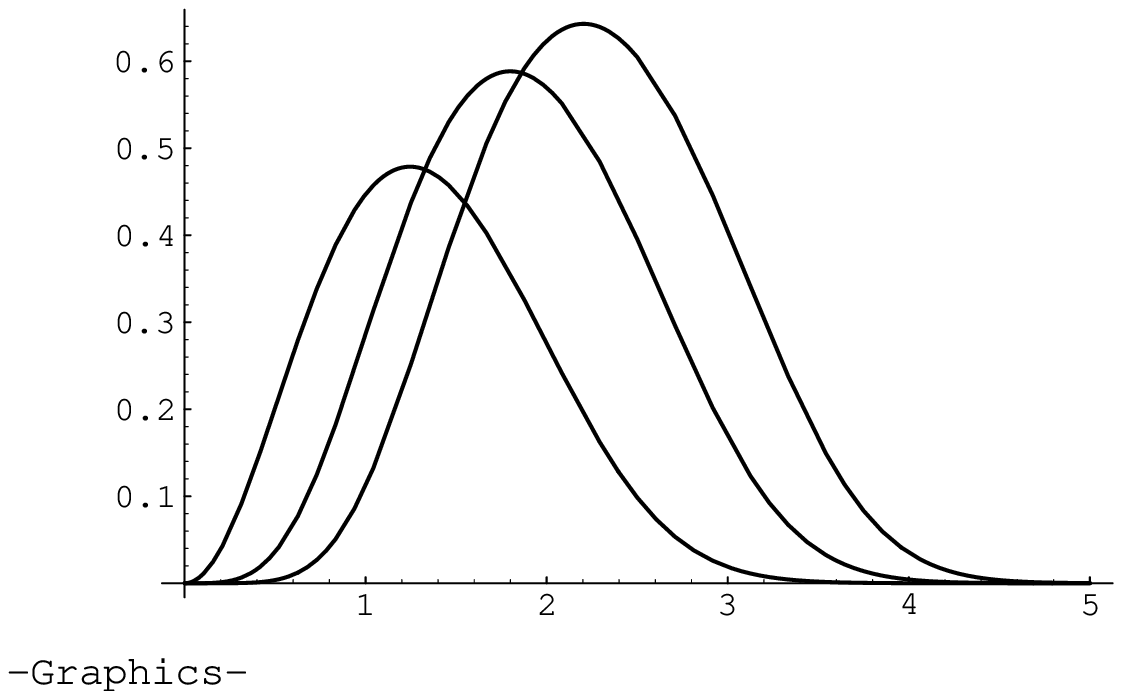}}
\vskip-0.5truecm

\kikezd{Fig. 2.}\hskip2mm
{\it The order parameter density of the 
radially symmetric non-topological vortices
for $n=1, 2, 3$. For each value of $n$, there is a full 
range of vortices}.
\goodbreak
\vskip1mm
\goodbreak

The arisal of these two different types of vortex solutions 
can be understood by
 adding a self-interaction potential, 
$U(\psi)$, to the Lagrangian. The use of such a potential
is quite commun in
Landau-Ginzburg theory, see Ref. [LP], p. 179.
In the context of the Quantum Hall Effect,
it can be viewed as the remnant of the
two-body potential in the second-quantized
Hamiltonian for spin-polarized electrons, when the
effective theory is derived [ZH].
\goodbreak
$\bullet$ For the symmetry breaking potential
$$
U(\psi) = {\lambda \over 8} (1-|\psi|^2)^2,
\equation
$$
finite energy requires $|\psi| \to 1$ at infinity.
Then the Bogomolny Eq.~(3) yields topological vortex 
solutions, provided
$$
\lambda=-{4\over\kappa^2}\pm{8\over\kappa}.
\equation
$$
The potential in (14) is physically admissible
(repulsive)
when $\lambda\geq0$, i.e. $|\kappa|\geq1/2$.

$\bullet$ For the 
non-symmetry-breaking potential
$$
\overline{U}=C+{\lambda\over 8}|\psi|^4
\equation
$$
finite-energy requires instead $|\psi| \to 0$ at infinity, and the 
Bogomolny trick works
when $C=-{1/2\kappa^2}$
and for $\lambda$ as in Eq. (15). 
The potential (16) is physically admissible (attractive)
when $\lambda\leq0$, i.e.,
when $\vert\kappa\vert\leq1/2$.
Changing $\lambda$ from a negative to a positive value
can be viewed as a phase transition, with transition point
$\lambda=0$. 

Let us stress that there is no way to obtain the
quantization of the Hall conductance from our classical field theory~:
the condition (4) is merely replaced by Eqn. (15).

Thus, the modified Girvin model (1) can be viewed as the $\lambda\to 0$ 
limit of {\it two}, slightly different systems, one of them
correct for $\kappa\geq1/2$, the other for $\kappa\leq1/2$.
In both cases, there is a natural boundary
condition at infinity, dictated by finite-energy. 
However,  $|\kappa|\to 1/2$ when $\lambda\to 0$,
so that both conditions can be used.
In this limit,
the boundary condition at infinity
 has to be put in by hand, 
since one can not know what 
stays `behind' the coefficient $\lambda=0$.  
(This happens also for 't 
Hooft--Polyakov monopoles
in the Prasad--Sommerfield limit). 

In Ref. [HHY], we have shown that, for the potential (14) with
$\lambda\neq0$, the system admits a 6-parameter group of symmetry,
made of the unbroken parts of the ``geometric'' and
``hidden'' Schr\"odiger symmetries. It is easy to see that, for
$\lambda=0$ some of the obstructions are lifted so that
``imported'' dilatations and expansions are also unbroken.
Thus, we have the
full ``hidden'' Schr\"odiger symmetry, just like for the
purely quartic potential [EZA], [JPH].

It is worth pointing out that the field-theoretical 
generalization of the
Girvin model is the {\it non-relativistic Maxwell--Chern--Simons} 
system proposed by
Manton [MAN], whose $2+1$-dimensional Lagrangian reads
$$
{1\over2}b^2-{i\over2}\big(\psi^*D_t\psi-\psi(D_t\psi)^*\big)
+ 
|\vec{D}\psi|^2
+
U(\psi)-{\kappa\over2}(ba_t+\vec{a}\times\vec{e}) 
+
a_t+\vec{a}\cdot{\vec{\jmath}}\strut{}^T,
\equation
$$
where the constant vector ${\vec{\jmath}}\strut{}^T$ 
(called the transport current) has been included
for the sake of Galilean invariance. 
Note also the absence of an electric term.
In a suitable Galilei frame ${\vec{\jmath}\strut}{}^T$ vanishes 
[MAN] and, for $U(\psi)\equiv0$ and identifying
$a_t$ with $-i\phi$, the modified Girvin system (1) is recovered,
when time-independence is assumed. 

So far, we have been working with a spinless matter field.
Introducing spin would not change the situation, though.
Modifying the Lagrangian (1) as
\eq=1
$$
{\cal L}_{spin}=
{1\over2}b^2+(\vec{D}\Psi)^\dagger(\vec{D}\Psi)
+i\phi\big(\Psi^\dagger\Psi-1\big)
-b\,
\Psi^\dagger\sigma_3\Psi
-i{\kappa\over2}\big(b\phi+\vec{a}\times\vn\phi\big)
\eqletter{''}
$$
where $\Psi$ is a $2$-component Pauli spinor,
would replace (2a) by the Pauli equation
\eq=2
$$
i\phi\Psi=\big[\vec{D}^2+b\sigma_3\big]\Psi,
\eqletter{a''}
$$
while (2b) and (2c) remain unchanged up to
$\vec\jmath=-i(
\Psi^\dagger\vec{D}\Psi-(\vec{D}\Psi)^\dagger\Psi)
+\vn\times(\Psi^\dagger\sigma_3\Psi).
$ 
Then, for $\kappa=\pm(1/2)$, the spinning system admits 
self-dual solutions of definite chirality, 
$\Psi_+=\pmatrix{0\cr\psi_+}$
and
$\Psi_-=\pmatrix{\psi_-\cr0}$,
with $\varrho=|\Psi_\pm|$ satisfying the {\it same}
Liouville-type equation (5) [HHY].

In Ref. [ZH], Zhang {\it et al.\/} proposed 
another `Landau-Ginzburg' theory for the QHE.
They consider a scalar field $\psi$ coupled to a gauge field 
$A_{\mu}$,
described by the Lagrangian
\eq=17
$$\eqalign{
\L_{Z}\;=\;&4\theta\,\epsilon^{ij}
\big(2A_0\p_{i}A_{j}-A_{i}\p_{0}A_{j}\big)
-{1\over4\theta}\epsilon^{\mu\nu\sigma}A_{\mu}\p_{\nu}A_{\sigma}
\ccr
&
+\psi^*\big[i\p_{0}-(A_{0}+A_{0}^{ext})\big]\psi
+\psi^*\big[-i\vn-(\vec{A}+\vec{A}^{ext})\big]^{2}\psi
+U(\psi).
\cr}
\equation
$$
where $A_{\mu}^{ext}$ is the vector potential of
an external electromagnetic field, and
$U(\psi)=\mu\vert\psi\vert^2-\lambda\vert\psi\vert^{4}$
is a quartic self-interaction potential.
They argue that their theory
is different from that of Girvin and MacDonald.
Now we show that for a static system in a purely magnetic 
background and
for $U(\psi)\equiv0$, the two models are indeed
{\it mathematically equivalent}.
To see this, let us note that under the above restrictions,
after some
partial integrations and dropping surface terms,
the Lagrangian of Zhang {\it et al.\/} becomes
$$
\Big(4\theta-{1\over4\theta}\Big)
\epsilon^{\mu\nu\sigma}A_{\mu}\p_{\nu}A_{\sigma}
-A_{0}\vert\psi\vert^{2}
+\big\vert(-i\vn-(\vec{A}+\vec{A}^{ext})\psi\big\vert^{2}.
\equation
$$

On the other hand, 
the Girvin-MacDonald model can also be presented in
a slightly different way.
Let us indeed consider a static, purely 
magnetic external field $B^{ext}$.
Then, setting
$
\vec{a}=-\vec{A}-\vec{A}^{ext}
$
and
$
A_{0}=-i\phi,
$
we find that, for the choice 
$$
\kappa={1\over B^{ext}},
\equation
$$
 the 
(original) Girvin-MacDonald Lagrangian [i.e., (1) {\it without} the 
$b^{2}/2$ term] becomes, up to a surface term,
$$
\L_{G}=\big\vert\vn-i(\vec{A}+\vec{A}^{ext})\psi\big\vert^{2}
-A_{0}\vert\psi\vert^2
-{\kappa\over2}\Big(A_{0}\vn\times\vec{A}+\vec{A}\times\vn A_{0}\Big),
\equation
$$
which is indeed (18) when
$\kappa=-8\theta+1/2\theta$\foot{
In the reprint volume
of M.~Stone: {\sl Quantum Hall Effect}, 
(Singapore: World Scientific, 1992), the first term in Eq. (17) has 
been suppressed. This does not change the conclusion~:
the two models are still equivalent when $\kappa=1/2\theta$.}.

In Ref. [EZA] Ezawa {\it et al.\/} have shown that the
model of Zhang {\it et al.\/} admits, for a suitable
choice of the self-interaction potential $U(\psi)$,
topological as well as non-topological vortex solutions. 
In the light of our results we see that, alternatively, 
we can add a $b^{2}$ term to the Lagrangian while
still working with $U(\psi)=0$.

As to the physical significance of our solutions, our ``topological'' 
vortices correspond to quasiparticles and quasiholes.
The physical interpretation of our ``non-topological'' 
vortices is, however, not yet clear. We are currently working on this 
issue.

\goodbreak
\kikezd{Acknowledgements}.
 M.~H. and J.-C.~Y. acknowledge
the {\it Laboratoire de Math\'emathiques et de Physique Th\'eorique}
of Tours University for hospitality.  
They are also indebted to the French Government
and  the Gouvernement de La C\^ote d'Ivoire
respectively, for doctoral scholarships.
P.~A.~H. would like to thank Drs. G.~Dunne and 
R.~Jackiw
for discussions and the hospitality in the Center for Theoretical Physics 
at MIT, where part of this work was completed.

\goodbreak
\vskip3mm\goodbreak
\centerline{\bf References}
\vskip-2mm

\reference[GIR]
S. M. Girvin, in {\sl The Quantum Hall Effect},
edited by R. E. Prange and S.~M.~Girvin (Springer Verlag,
N. Y. 1986), Chapt. 10.
See also S.~M.~Girvin and A.-H.~MacDonald, 
{\sl Phys. Rev. Lett}. {\bf 58}, 303 (1987).

\reference[LP]
E. M. Lifshitz and L. P. Pitaevski,
{\sl Statistical Physics}, Part 2
(Landau-Lifshitz Course of Theoretical Physics Vol. 9),
Butterworth-Heinemann: Oxford (1995).

\reference[MAN]
N. Manton, {\sl Ann. Phys}. (N.Y.) {\bf 256}, 114 (1997). 
Related results can also be found in
I. Barashenkov and A. Harin, {\sl Phys. Rev. Lett}. {\bf 72}, 1575 (1994);
{\sl Phys. Rev}. {\bf D52}, 2471 (1995);
 P. Donatis and R. Iengo, 
 {\sl Phys. Lett}. {\bf B320}, 64 (1994);
K. Lee and P. Yi, {\sl Phys. Rev}. {\bf D52}, 2412 (1995).

\goodbreak
\reference[JP]
R.~Jackiw and S-Y.~Pi, 
{\sl Prog. Theor. Phys. Suppl}. {\bf 107}, 1 (1992); 
G. Dunne, {\sl Self-Dual
Chern-Simons Theories}. Springer Lecture Notes in Physics. New Series: 
Monograph 36. (1995).

\reference[HHY]
M.~Hassa\"{\i}ne, P.~A.~Horv\'athy and 
J.-C.~Yera,
{\it Non-relativistic Maxwell--Chern--Simons vortices\/}, 
{\sl Annals of Physics} (N. Y.). {\bf 263}, 276 (1998).

\reference[BOG]
E. B. Bogomolny, 
{\sl Sov. J. Nucl. Phys}. {\bf 24}, 449 (1976);
H. J. De Vega and F. A. Schaposnik, 
{\sl Phys. Rev}. {\bf D14}, 1100 (1976).
\goodbreak

\reference  [ZH]
S. C. Zhang, T. H. Hansson, S. Kivelson,
{\sl Phys. Rev. Lett}. {\bf62}, 307 (1989).

\reference[EZA]
Z. F. Ezawa, M. Hotta and A. Iwazaki, {\sl Phys. Rev}. {\bf D44}, 452
(1991).

\reference[JLW]
R.~Jackiw, K.~Lee and E.~Weinberg, {\sl Phys. Rev}. {\bf D42}, 3488 
(1990).

\reference[WEIN]
E. Weinberg, {\sl Phys. Rev}. {\bf D19}, 3008 (1979);
C. H. Taubes,
{\sl Commun. Math. Phys}. {\bf 72}, 277 (1980).

\reference[JPH]
R.~Jackiw and S-Y.~Pi,
{\sl Phys. Rev. Lett}. {\bf 67}, 415 (1991);
{\sl Phys. Rev}. {\bf D44}, 2524 (1991);
M. Hotta {\sl Prog. Theor. Phys}. {\bf 86}, 1289 (1991).

\end